\documentclass[aps,pre,preprint,superscriptaddress]{revtex4-1}

\usepackage{graphicx}% Include figure files
\usepackage{bm}% bold math
\usepackage{color}
\usepackage{amsmath}
\usepackage{amssymb}
\usepackage{mathtools}
\usepackage{dsfont}
\usepackage{color}
\usepackage{calc}
\usepackage{hyperref}
\usepackage{multirow}
\usepackage{natmove}
\usepackage{epstopdf}%This line makes .eps figures into .pdf - please comment out if not required.
\newcommand{\lla}{\left\langle}
\newcommand{\rra}{\right\rangle}

\newcommand{\bs}[1]{\boldsymbol{#1}}

\newcommand{\vphi}{\varphi}

\graphicspath{{./figures/}}

\begin{document}
\title{Conformational properties of active semiflexible polymers}

\author{Thomas Eisenstecken}
\email{t.eisenstecken@fz-juelich.de}
\affiliation{Theoretical Soft Matter and Biophysics, Institute for
Advanced Simulation and Institute of Complex Systems,
Forschungszentrum J\"ulich, D-52425 J\"ulich, Germany}
\author{Gerhard Gompper}
\email{g.gompper@fz-juelich.de}
\affiliation{Theoretical Soft Matter and Biophysics, Institute for
Advanced Simulation and Institute of Complex Systems,
Forschungszentrum J\"ulich, D-52425 J\"ulich, Germany}
\author{Roland G. Winkler}
\email{r.winkler@fz-juelich.de}
\affiliation{Theoretical Soft Matter and Biophysics, Institute for
Advanced Simulation and Institute of Complex Systems,
Forschungszentrum J\"ulich, D-52425 J\"ulich, Germany}

%\thanks{}

\date{\today}

\begin{abstract}
The conformational properties of flexible and semiflexible polymers exposed to active noise are studied theoretically. The noise may originate from the interaction of the polymer with surrounding active (Brownian) particles or from the inherent motion of the polymer itself, which may be composed of active Brownian particles. In the latter case, the respective monomers are independently propelled in directions changing diffusively. For the description of the polymer, we adopt the continuous Gaussian semiflexible polymer model. Specifically, the finite polymer extensibility is taken into account which turns out to be essentially for the polymer conformations. Our analytical calculations predict a strong dependence of the relaxation times on the activity. In particular, semiflexible polymers exhibit a crossover from a bending-elasticity-dominated to the flexible-polymer dynamics with increasing activity. This leads to a significant noise-induced polymer shrinkage over a large range of self-propulsion velocities. For large activities, the polymers swell and their extension  becomes comparable to the contour length. The scaling properties of the mean square end-to-end distance with respect to the polymer length and monomer activity are discussed.
{\let\newpage\relax}
\end{abstract}
\pacs{}
\keywords{}
%{\let\newpage\relax\maketitle}

%\maketitle

\begingroup
\let\newpage\relax%
\maketitle
\endgroup

\section{Introduction}

A distinctive characteristics of active matter is the conversion of internal chemical energy into, or utilization of energy from the environment for, directed motion \cite{laug:09,rama:10,vics:12,roma:12,marc:13,elge:15,bech:16,marc:16.1,zoet:16}.
The spectrum  of biological active systems is wide and ranges from the the macroscopic scale of flocks of birds and mammalian herds \cite{vics:12}, the cytoskeleton in living cells \cite{nedl:97,howa:01,krus:04,baus:06,juel:07,hara:87,scha:10,rama:10,marc:13,pros:15}, down to  moving bacteria \cite{berg:04,elge:15,rama:10} on the micrometer scale. Thereby,  nature employs various propulsion strategies. Bacteria are typically propelled by helical flagella \cite{elge:15,berg:04,scha:02,cope:09,kear:10}. The actin filaments of the cytoskeleton are driven forward by molecular motors \cite{hara:87,scha:10,juel:07,marc:13,pros:15,cord:14}. Alike,  microtubules in motility assays are propelled  by surface-bound dyneins  \cite{sumi:12}. For synthetic active particles, chemical or physical propulsion mechanism are exploited \cite{hows:07,volp:11,butt:13,hage:14}.

Various features are common to all active systems \cite{wink:16}, and the challenge of a theoretical description is to find a suitable approach capturing these characteristics. Generically, the activity-induced hydrodynamic flow field of a microswimmer is described by a force dipole \cite{kim:13,laug:09,dres:11}. Experiments, theoretical calculations, and computer simulations, e.g.,  for {\em E. coli} bacteria \cite{dres:11,dres:10.1,guas:10,wata:10,hu:15.1} and {\em Chlamydomonas
reinhardtii} algae \cite{dres:10.1,guas:10,ghos:14.1,klin:15}, confirm such a description for the far-field flow.
However, the near-field flow can be distinctively different from the flow field of a force dipole \cite{hu:15.1,dres:10.1,guas:10,ghos:14.1,klin:15}.

Microswimmers are often described as active Brownian particles (ABPs) \cite{hows:07,peru:10,roma:12,fily:12,bial:12,redn:13,wyso:14,wink:16,hage:15}, neglecting hydrodynamics. This minimal stochastic
model already yields interesting propulsion and excluded-volume induced emerging structures \cite{roma:12,fily:12,bial:12,redn:13,wyso:14}. Moreover, ABPs are an extremely useful  model to unravel
the out-of-equilibrium statistical features of active
systems \cite{yang:14,solo:15,solo:15.1,taka:14,magg:15,gino:15,bert:15,spec:15,wink:15}.

The properties of connected active particles, such as linear chains
\cite{live:01,sark:14,chel:13,loi:11,ghos:14,isel:15,isel:16,lask:13,jaya:12,jian:14,babe:16,kais:14,vale:11,suma:14,cugl:15,wink:16} or other arrangements \cite{kuch:16}, are particular interesting systems, because of the coupling of their conformational properties and propulsion. Similar to external forces, the intrinsic activity leads to significant conformational changes, as shown in Refs.~\cite{kais:15,isel:15,wink:16}. In this context, we also like to mention the conformational modulations of polymer embedded in a bath of active Brownian particles \cite{hard:14,shin:15}. Activity also affects other polymer properties. An example is the linear viscoelastic response of an entangled,
isotropic solution of semiflexible polymers as a model systems for myosin-driven actin filaments \cite{live:01}. Here, activity  leads to novel time-dependent regimes of the shear modulus. Other aspects are emerging beat patterns \cite{chel:13}, activity-induced ring closure \cite{sark:14,sama:16}, aggregation of individual polymers in two dimensions \cite{isel:15},  and collective phenomena \cite{loi:11}. Moreover, the internal dynamics of active dumbbells \cite{wink:16} and polymers \cite{ghos:14,sama:16} has been addressed.  The influence of hydrodynamic interactions on the dynamical properties of active polymer properties have been analyzed in Refs.~\cite{jaya:12,lask:13,lask:15,babe:16}.

The (theoretical) analysis of the nonequilibrium behavior of flexible and semiflexible polymers, e.g., under shear flow
\cite{dua:00,prab:06,dua:00.1,wink:06,munk:06,wink:10} or during stretching \cite{bird:87,wink:92,mark:95,wink:03,wink:10.1,kier:04,salo:06,blun:09,lamu:12,hsu:12,radh:12,manc:12,manc:13,ilia:14,alex:15},
reveals the paramount importance of the finite polymer extensibility. We expect this intrinsic polymer property to be essential also for polymers comprising active monomers. Most theoretical  studies have neglected finite polymer extensibility \cite{kais:15,ghos:14,sama:16}. Only in the analytical treatment of the dynamics of an active dumbbell in Ref.~\cite{wink:16}, the finite extensibility has been taken into account and its fundamental importance for  the dumbbell dynamics has been demonstrated.

In this article, the conformational properties of flexible and semiflexible active Brownian polymers (ABPO) are studied analytically. Thereby, we consider a polymer composed of active Brownian particles, which are assembled in a linear chain. The diffusive motion of the propulsion velocity of the monomers is described by a Gaussian but non-Markovian process. The emphasize is on the conformational properties due to the intimate coupling of the entropic polymer degrees of freedom and the activity of the monomers. We adopt the Gaussian semiflexible polymer model \cite{wink:94,wink:03}, which allows us to treat the problem analytically. As an important extension to previous studies, we account for the finite polymer extensibility and demonstrate that it strongly affects the out-of-equilibrium properties of an active polymer. Evaluation of the polymer relaxation times shows a drastic influence of that constraint on the polymer dynamics.  In general, the relaxation times decrease with increasing activity, whereby the decline is more pronounced for stiffer polymers. Here, activity induces a transition from semiflexible polymer behavior, determined by bending elasticity, to entropy-dominated behavior of flexible polymers with increasing activity. Correspondingly, the conformational properties depend on activity.  In the simpler case of flexible polymers, activity leads to their swelling over a wide range of activities. Thereby, the dependence on activity is very different from the  theoretical prediction of a Rouse model \cite{kais:15}.  Interestingly, semiflexible polymers exhibit an activity induced shrinkage. However, for large activities the polymer conformations are ultimately comparable with those of flexible polymers.  Shrinkage of active polymers in two dimensions has been observed by simulations in Ref.~\cite{kais:15}. However, that shrinkage is due to excluded-volume effects and is unrelated to our observations for semiflexible polymers, where excluded-volume interactions are negligible.

Our theoretical considerations shed light on the nonequilibrium properties of semiflexible polymers and underline the importance of an adequate description already for moderate activities. Models without the constraint of a finite contour length, e.g., the standard Rouse model \cite{doi:86}, would by no means be able to reproduce and capture the correct structural and dynamical aspects.

\section{Model of Active Polymer} \label{sec:model}

\begin{figure}[t]
\begin{center}
\includegraphics*[width=0.7\columnwidth]{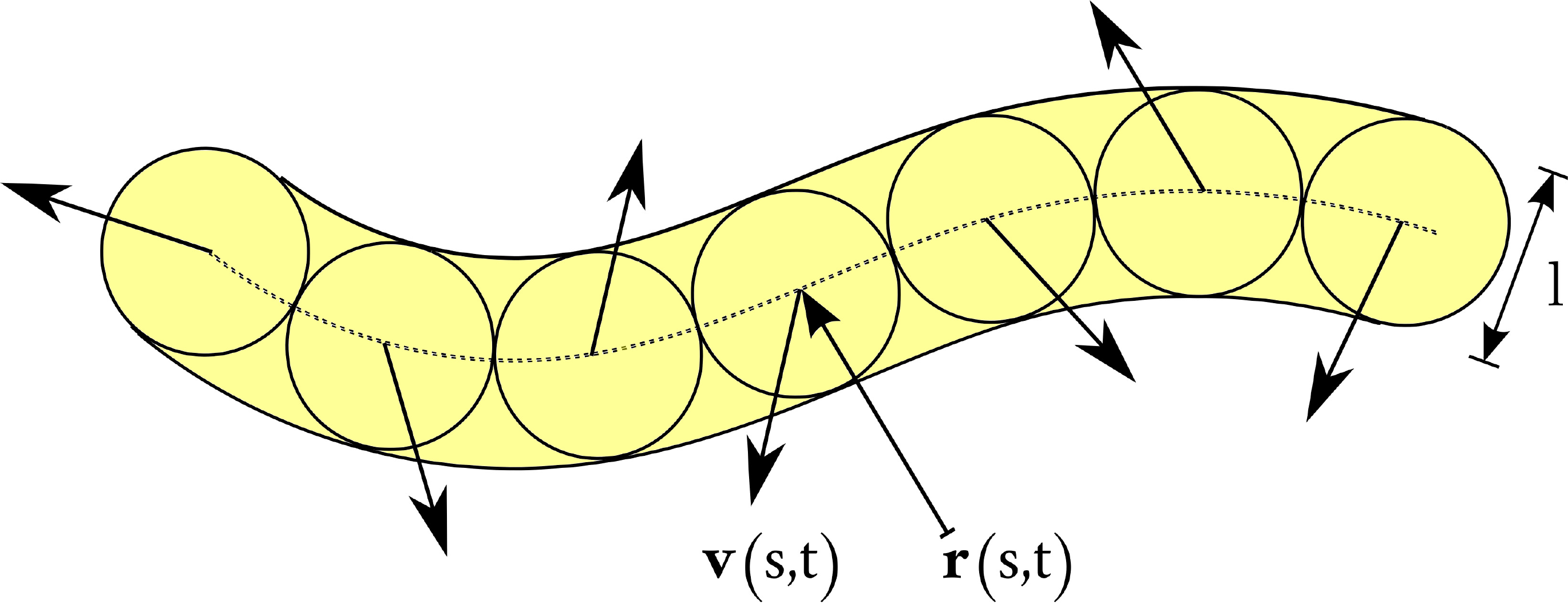}
\end{center}
\caption{Model of the continuous semiflexible active polymer.} \label{fig:model}
\end{figure}

We adopt a mean-field model for a semiflexible polymer \cite{bawe:85,batt:87,lang:91,wink:94,ha:95,wink:03}, which is
denoted as Gaussian semiflexible polymer (GSFP), complemented by the activity of the monomers (GSFAP). We describe the GSFP as a continuous, differentiable space curve ${\bm r}(s,t)$, where $s$ ($-L/2 \le s \le L/2$) is the contour coordinate along the chain of length $L$ and $t$ is the time. Activity is added by assigning the self-propulsion velocity  $\bm v(s,t)$ to every point $\bm r(s,t)$, as typical for active Brownian particles (cf. Fig.~\ref{fig:model}) \cite{elge:15,bial:12,fily:12,wyso:14,marc:16.1,bech:16}.  The equation of motion of the GSFAP is then given by the Langevin equation
\cite{harn:95,wink:97,wink:10,wink:06.1,wink:07.1}
\begin{align} \label{eq:equation_motion}
\frac{\partial}{\partial t}{ \bm r}(s,t) = \bm v(s,t) + \frac{1}{\gamma} \left( 2 \lambda k_B T \frac{\partial^2}{\partial s^2} { \bm r}(s,t)  - \epsilon k_B T
\frac{\partial^4} {\partial s^4} { \bm r}(s,t) + {\bm \Gamma} (s,t) \right) \:,
\end{align}
with the boundary conditions
\begin{align} \label{eq:boundary_conditions_1}
\left[2 \lambda \frac{\partial}{\partial s} {\bm r}(s,t) - \epsilon
\frac{\partial^3}{\partial s^3} {\bm r}(s,t) \right]_{\pm L/2}  =  0 \ , \hspace{3mm} \left[2 \lambda_0 \frac{\partial}{\partial s} {\bm
r}(s,t) \pm \epsilon \frac{\partial^2}{\partial s^2} {\bm r}(s,t) \right]_{\pm
L/2} =  0 \:.
\end{align}
The terms with the second and fourth derivative in Eq.~(\ref{eq:equation_motion}) account for the entropic degrees of freedom and bending restrictions, respectively. Formally, the entropic part looks like a stretching energy due to harmonic bonds along the polymer contour with  $\lambda k_BT$ and $\lambda_0 k_BT$ as the Hookean spring constants  \cite{bird:87,oett:96} of the continuous chain. In the following, we will denote $\lambda$ and $\lambda_0$ as stretching and $\epsilon$ as bending coefficient. Note that $\lambda$ and $\lambda_0$ are in general different due to the broken symmetry at the chain ends.
The stochastic force ${\bm \Gamma} (s,t)$ is assumed to be stationary, Markovian, and Gaussian with zero mean
and the second moments
\begin{align} \label{eq:noise_translation}
\lla \Gamma_{\alpha} (s,t) \Gamma_{\beta}(s',t') \rra = 2 \gamma  k_B T \delta_{\alpha \beta} \delta(s-s') \delta(t-t') \ ,
\end{align}
where $T$ is the temperature,  $k_B$ the Boltzmann constant, $\gamma$ the translational friction coefficient per length, and $\alpha, \beta \in\{x,y,z\} $. The Lagrangian multipliers $ \lambda$, $\lambda_0$, and
$\epsilon$ are determined by constraints \cite{wink:92,wink:03}. In general, we  find  $\epsilon =3/4p$ and $\lambda_0=3/4$ for a polymer in three dimensions, where $p$ is related to the persistence
length $l_p$ via $p=1/2l_p$ \cite{wink:92,wink:03}, i.e., the bending coefficient $\epsilon=3l_p/2$ is solely determined by the persistence length as is well known \cite{krat:49,arag:85,wink:07.1}. In Eq.~(\ref{eq:equation_motion}), we apply a mean-field value for the Lagrangian multiplier $\lambda$. Strictly, we expect the Lagrangian multiplier to depend on the contour coordinate for the active system, because,  as shown in Refs.~\cite{wink:92,wink:03,wink:06,wink:10,wink:10.1}, $\lambda$ strongly depends on the presence of an external force, i.e., $\lambda = \lambda(s)$, since it is determined by the local inextensibility condition $\lla (\partial {\bm r} / \partial s)^2 \rra =1$.  However, in Eq.~(\ref{eq:equation_motion}), we neglected this aspect and assume that $\lambda$ is constant along the polymer contour.  Hence, we imply the global constraint of a finite contour length
\begin{align} \label{eq:constraint}
\int_{-L/2}^{L/2} \lla \left(\frac{\partial \bm r (s,t)}{\partial s}\right)^2 \rra d s = L
\end{align}
corresponding to a mean-field approach.
As a consequence, the polymer conformations may be inhomogeneous along its contour as, e.g., in the stretching of the GSFP \cite{wink:03}. However,
the full solution of a discrete free-draining polymer model with
individual Lagrangian multipliers for every bond and bond angle
\cite{wink:92,wink:94,wink:03},  yields expectation
values for global quantities such as viscosity which deviate only
very little from those determined with the constraint
(\ref{eq:constraint}) in the limit of a nearly continuous polymer.
Hence, the solution of the equations of motion with the constraint
(\ref{eq:constraint}) suffices for many practical purposes.

We regard the self-propulsion velocity $\bm v(s,t)$ as a non-Markovian stochastic process in time with the correlation function
\begin{align} \label{eq:corr_colored}
\lla \bm v(s,t) \cdot \bm v(s',t')\rra = v_0^2 l e^{-\gamma_R(t-t')} \delta(s-s') \ .
\end{align}
Here,  $v_0$ the magnitude of the propulsion velocity and $\gamma_R$ the damping factor of the rotational motion.
The velocity correlation function arises, on the one hand, from the independent stochastic process for the propulsion velocity
\begin{align}
\frac{\partial}{\partial t} {\bm v}(s,t) = - \gamma_R \bm v(s,t) + \bs{\eta} (s,t) ,
\end{align}
where $\bs{\eta}(s,t)$ is a Gaussian and Markovian stochastic forces with zero mean and the second moment
\begin{align} \label{eq:noise_velocity}
\lla \bs{\eta}(s,t) \cdot \bs{\eta} (s',t') \rra = 4 D_R v^2_0 l \delta(s-s') \delta(t-t')
\end{align}
in three dimensions; $D_R=\gamma_R/2$ is rotational diffusion coefficient. On the other hand, the  correlation function (\ref{eq:corr_colored}) also follows for the active force $\gamma v_0 \bm e(s,t)$, with a constant self-propulsion velocity $v_0$ and  the unit vector
$\bm e$  of the propulsion direction, where $\bm e$  performs a random walk according to \cite{elge:15,wink:15,wink:16,marc:16.1}
\begin{align} \label{eq:orient}
\frac{\partial}{\partial t} \bm e(s,t) = \hat{\bm \eta}(s,t) \times \bm e(s,t) .
\end{align}
Here, $\hat{\bm \eta}(s,t)$ is a Gaussian and Markovian stochastic process with zero mean and the second moment
\begin{align} \label{eq:noise_velocity_orient}
\lla \hat{\bm \eta}(s,t) \cdot \hat{\bm \eta}(s,t) \rra = 4 D_R  l \delta(s-s') \delta(t-t') .
\end{align}
Since we will need and apply only the correlation function (\ref{eq:corr_colored}) in the following, the exact nature of the underlying process is irrelevant and our considerations apply for both type of processes.

Note that the continuum representation of the semiflexible polymer requires to introduce a length scale $l$ in Eqs.~(\ref{eq:corr_colored}) and (\ref{eq:noise_velocity}).  With a touching-bead model in mind for a discrete polymer, this minimum length corresponds to the bead diameter and bond length of that model (cf. Fig.~\ref{fig:model}). Strictly speaking, $l$ is a free parameter in the continuum model. For a flexible polymer, we regard $l=2l_p=1/p$ as the Kuhn length \cite{flor:89,rubi:03}.

In the above description, we consider the velocity $\bm v$ as an intrinsic property of the active polymer. However, we may also consider $\bm v$ as an external stochastic process with an exponential correlation (colored noise) \cite{elge:15,wink:16,sama:16,marc:16.1}. Such a correlated noise may be exerted by active Brownian particles on an embedded polymer \cite{shin:15,hard:14,kais:14}.

\section{Solution of Equation of Motion} \label{sec:solution_eom}

To solve the equation of motion (\ref{eq:equation_motion}), we apply an eigenfunction expansion in terms of
the eigenfunctions of the eigenvalue equation \cite{wink:06,harn:95}
\begin{eqnarray} \label{eigenvalue_equation}
\epsilon  k_B T \frac{d^4}{d s^4} \vphi_n(s) - 2  \lambda k_B T \frac{d^2}{d s^2 }
\vphi_n(s) = \xi_n \vphi_n(s)\:.
\end{eqnarray}
The resulting eigenfunctions are given by \cite{wink:06,harn:95}
\begin{align} \label{eigenfunctions}
\vphi_0  = & \sqrt{\frac{1}{L}}, \\ %%
\vphi_n(s)  = & \sqrt{\frac{c_n}{L}}\left(\zeta_n'\frac{\sinh \zeta_n' s}{ \cosh
\zeta_n' L/2} + \zeta_n \frac{\sin \zeta_n s}{ \cos \zeta_nL/2} \right) ,
\ n \ \mbox{odd} ,\\  %%
\vphi_n(s)  = & \sqrt{\frac{c_n}{L}} \left(\zeta_n'\frac{\cosh \zeta_n' s}{
\sinh \zeta_n' L/2} -\zeta_n \frac{\cos \zeta_n s}{ \sin \zeta_n L/2} \right) ,
\ n \ \mbox{even} \:,
\end{align}
with
\begin{align}
\zeta_n'^2 - \zeta_n^2 = \frac{2 \lambda}{\epsilon} \ , \ \xi_0 =0 \ , \ \xi_n = k_BT(\epsilon \zeta_n^4 + 2 \lambda \zeta_n^2) \ .
\end{align}
The $c_n$s follow from the normalization condition, and the wave numbers $\zeta_n$
and $\zeta_n'$ are determined by the
boundary conditions (\ref{eq:boundary_conditions_1}). $\vphi_0$ describes the translational motion of
the whole molecule.

Inserting the eigenfunction expansions
\begin{equation} \label{eq:eigen_expansion}
\begin{split}
{\bm r}(s,t) = \sum_{n=0}^{\infty} \bs \chi_n(t) \vphi_n(s) , \: \: \bs \Gamma
(s,t) = \sum_{n=0}^{\infty} \bs \Gamma_n(t) \vphi_n(s) , \\ \: \: \bs \eta
(s,t) = \sum_{n=0}^{\infty} \bs \eta_n(t) \vphi_n(s) , \: \: \bs
{\bm v}(s,t) = \sum_{n=0}^{\infty} \bm v_n(t) \vphi_n(s)
\end{split}
\end{equation}
into Eq.~(\ref{eq:equation_motion}) yields the equation of motion for the mode amplitudes $\bs \chi_n$
\begin{equation} \label{eq:equation_amplitudes}
\frac{d}{d t}\bs \chi_n(t) = - \frac{1}{\tau_n} \bs{\chi}_n (t) + \bm v_n(t)  +
\frac{1}{\gamma}  \bs \Gamma_n (t)  \: ,
\end{equation}
with the relaxation times
\begin{align} \label{eq:relax_time}
\tau_n = \frac{\gamma}{\xi_n} = \frac{\gamma}{ k_BT(\epsilon \zeta_n^4 + 2 \lambda \zeta_n^2) } \ .
\end{align}
The stationary-state solution of  Eq.~(\ref{eq:equation_amplitudes}) is
\begin{align}
\bs \chi_n(t) = e^{-t/\tau_n} \int_{-\infty}^t e^{t'/\tau_n} \left( \bm v_n(t') + \frac{1}{\gamma} \bs \Gamma_n (t') \right) dt' \ .
\end{align}
The time correlation functions of the mode amplitudes, which are useful in the further analysis, are obtained as $\lla \bs \chi_n(t) \cdot \bs \chi_m(t')\rra = \delta_{nm} \lla \bs \chi_n(t) \cdot \bs \chi_n(t')\rra$, with \cite{wink:16}
\begin{align} \label{eq:correlation}
\lla \bs \chi_n(t) \cdot \bs \chi_n(t')\rra =   \left(  \frac{3 k_BT \tau_n}{\gamma} e^{-|t-t'|/\tau_n}  + \frac{v_0^2 l \tau_n ^2}{1-(\gamma_R \tau_n)^2} \left[e^{-\gamma_R |t-t'|} - \gamma_R \tau_n e^{-|t-t'|/\tau_n} \right] \right) \ .
\end{align}

\section{Results}

\subsection{Center-of-Mass Motion}

The center-of-mass position is given by \cite{harn:95,wink:06.1}
\begin{align}
\bm r_{cm} (t) = \frac{1}{L} \int_{-L/2}^{L/2} \bm r(s,t) \, ds = \bs \chi_0(t) \vphi_0(t) \ .
\end{align}
With the solution of Eq.~(\ref{eq:equation_amplitudes}) for the zeroth's mode
\begin{align}
\bs \chi_0(t) = \bs \chi_0(0) + \int_0^t \left( \bm v_n(t')  +
\frac{1}{\gamma}  \bs \Gamma_n (t') \right)   dt'   \ ,
\end{align}
we obtain the center-of-mass mean square displacement
\begin{align}
\lla \left( \bm r_{cm} (t) - \bm r_{cm} (0)\right)^2 \rra = \frac{6k_BT}{\gamma L} t + \frac{2 v_0^2 l}{\gamma_R^2L} \left(\gamma_R t - 1 + e^{-\gamma_R t} \right) \ .
\end{align}
As for an active Brownian particle, the term linear in time on the right-hand side accounts for the translational Brownian motion \cite{elge:15}. As a generalization, the total friction coefficient $\gamma L$ appears. The second term represents the contribution of activity. Again, it is similar to the term appearing for ABPs, aside from the ratio $L/l$. We can identify the latter as the number of frictional sites or monomers $N$ of diameter $l$, i.e., $L=Nl$. Then, $N=1$ corresponds to an ABP with the friction coefficient $\gamma l$, and $N=2$ to a dumbbell \cite{wink:16,comm1}.

The long-time diffusion coefficient follows as
\begin{align}
D = \frac{k_BT}{\gamma L} \left(1+\frac{3v_0^2 l \gamma}{\gamma_R k_BT} \right) = D_L \left(1 + \frac{3 Pe^2}{2 \Delta} \right) \ ,
\end{align}
with the diffusion coefficient $D_L=k_B T/\gamma L$ of a passive polymer, the P\'eclet number $Pe$, and the ratio $\Delta$ of the diffusion coefficients \cite{elge:15,sten:14,wink:16}
\begin{align} \label{eq:peclet}
Pe = \frac{v_0}{D_R l} \ , \ \ \ \Delta = \frac{D_T}{D_R l^2} \ .
\end{align}
Here, we introduce the diffusion coefficient $D_T = k_BT/\gamma l$ as the diffusion coefficient
of a segment of length $l$ (cf. description of the model on page 3). In the following, we use the thermal translational and rotational diffusion coefficients of spherical particles of diameter $l$ in solution, which yields $\Delta = 1/3$.

\subsection{Lagrangian Multiplier---Stretching Coefficient}

Inextensibility is a fundamental property of a polymer and determines its conformational and dynamical characteristics. Hence, we have to calculate the Lagrangian multiplier $\lambda$ first in order to relate other polymer aspects to the constraint Eq.~(\ref{eq:constraint}). Insertion of the eigenfunction expansion (\ref{eq:eigen_expansion}) for the position $\bm r(s,t)$  into Eq.~(\ref{eq:constraint}) yields
\begin{align} \label{eq:constraint_mode}
\sum_{n=1}^{\infty} \left( \frac{3 k_BT}{\gamma} \tau_n + \frac{v_0^2 l}{1+\gamma_R \tau_n} \tau_n^2 \right) \int_{-L/2}^{L/2} \left( \frac{d \vphi_n(s)}{ds} \right)^2 ds = L ,
\end{align}
which determines the Lagrangian multiplier $\lambda$. In terms of the P\'eclet number $Pe= v_0/D_Rl$ and $\Delta$ of Eq.~(\ref{eq:peclet}), this equation can be expressed as
\begin{align}
\sum_{n=1}^{\infty} \left( \frac{1}{\hat \xi_n} + \frac{Pe^2 N^3}{9 \Delta^2 \left(\hat \xi^2_n+ \frac{2N^3}{3 \Delta} \hat \xi_n \right)} \right) \int_{-1/2}^{1/2} \left( \frac{d \vphi_n(x)}{dx} \right)^2 dx = 1 \ ,
\end{align}
with the abbreviation
\begin{align}
\hat \xi_n = pL \mu (\zeta_nL)^2 + \frac{1}{4pL}(\zeta_nL)^4 \ .
\end{align}
Here, we introduce the Lagrangian multiplier $\mu$ via the relation $\lambda= 3 p \mu/2$, i.e.,  $\mu$ is the ratio between the stretching coefficients of the active and the passive polymer.  In the integral, we substituted $s$ by $x=s/L$.

\begin{figure}[t]
\begin{center}
\includegraphics*[width=0.7\columnwidth]{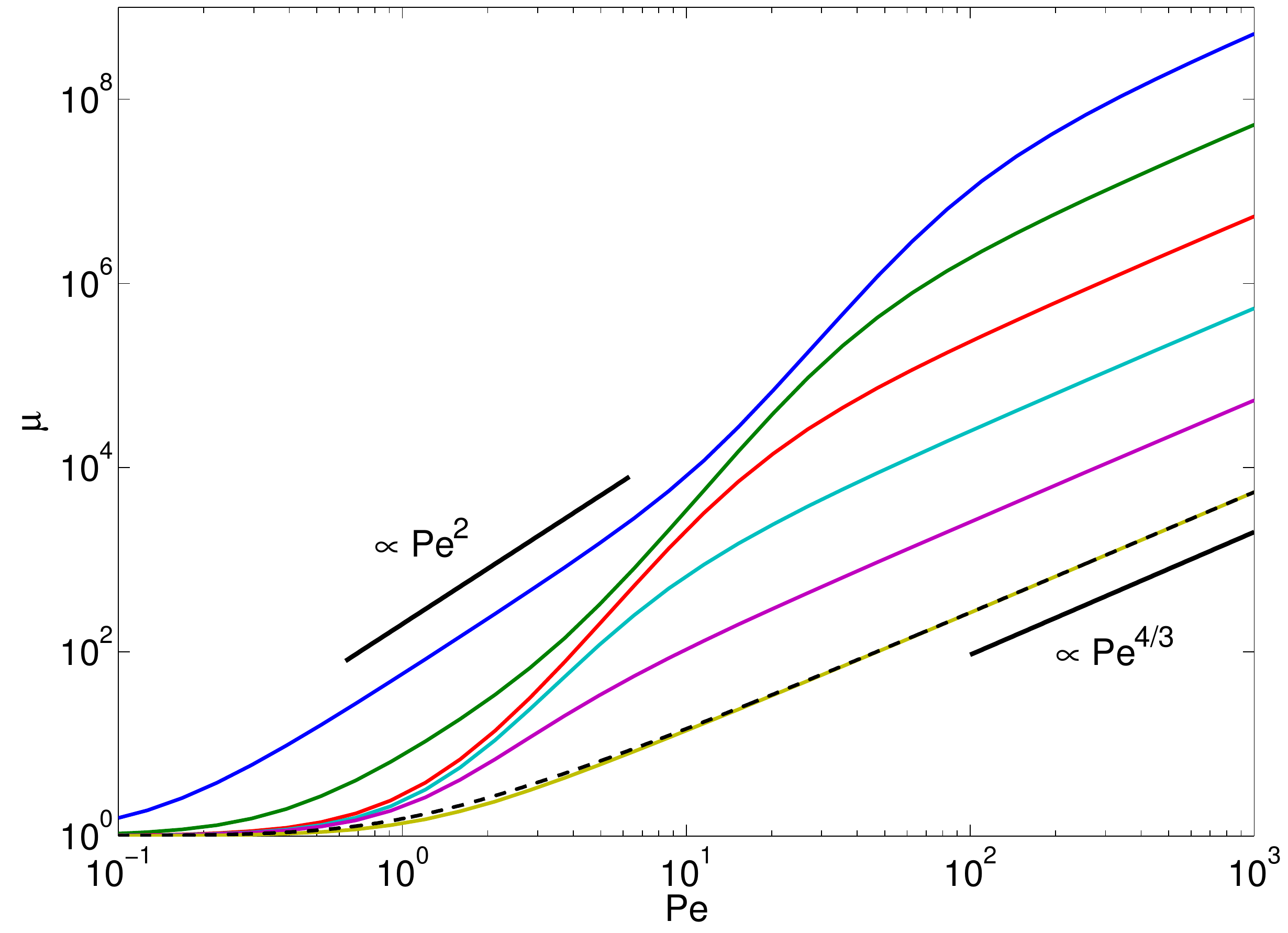}
\end{center}
\caption{Normalized stretching coefficient (Lagrangian multiplier) $\mu=2\lambda/3p$ as function of the P\'eclet number for the polymer bending stiffnesses $pL = 10^3$, $10^2$, $10$, $1$, $10^{-1}$, and $10^{-2}$ (bottom to top). For the other parameters, we set $N=L/l=10^3$ and $\Delta=1/3$. The dashed line for $pL=10^3$ represents the solution of the asymptotic equation (\ref{eq:lagpar_flex}). The straight lines indicate the power-law dependencies $\mu \sim Pe^{2}$ for $pL < 10^{-1}$ and $Pe<1$, and $\mu\sim Pe^{4/3}$ (cf. Eq.~(\ref{eq:limit_lagpar_pl})), respectively.} \label{fig:lagpar}
\end{figure}

Figure \ref{fig:lagpar} displays Lagrangian multipliers as function of the P\'eclet number for various bending stiffnesses $pL = L/2l_p$ (at constant polymer length $L$, variation of $pL$ corresponds to a variation of the polymer persistence length). Evidently, activity leads to an increase of the multiplier $\mu$ with increasing $Pe$. Thereby, semiflexible polymers with $pL \lesssim 10$ exhibit a pronounced dependence on $Pe$ already for moderate P\'eclet numbers. In the limit $Pe \to 0$, the multiplier assumes the value of a passive polymer $\mu=1$.  Over the considered range of P\'eclet numbers, the curves exhibit the asymptotic dependence $\mu \sim Pe^{4/3}$ for large $Pe$, independent of the polymer stiffness. For polymers with $pL \lesssim 10$, an intermediate regime appears, where $\mu \sim Pe^{\kappa}$, with $\kappa > 3$. Very stiff polymers ($pL< 10^{-1}$) even exhibit another power-law regime for small $Pe$, where $\mu \sim Pe^2$. The various activity-induced features reflected in the Lagrangian multiplier imply pronounced effects on the conformations and internal dynamics of an active polymer. \\

\begin{figure}[t]
\begin{center}
\includegraphics*[width=0.7\columnwidth]{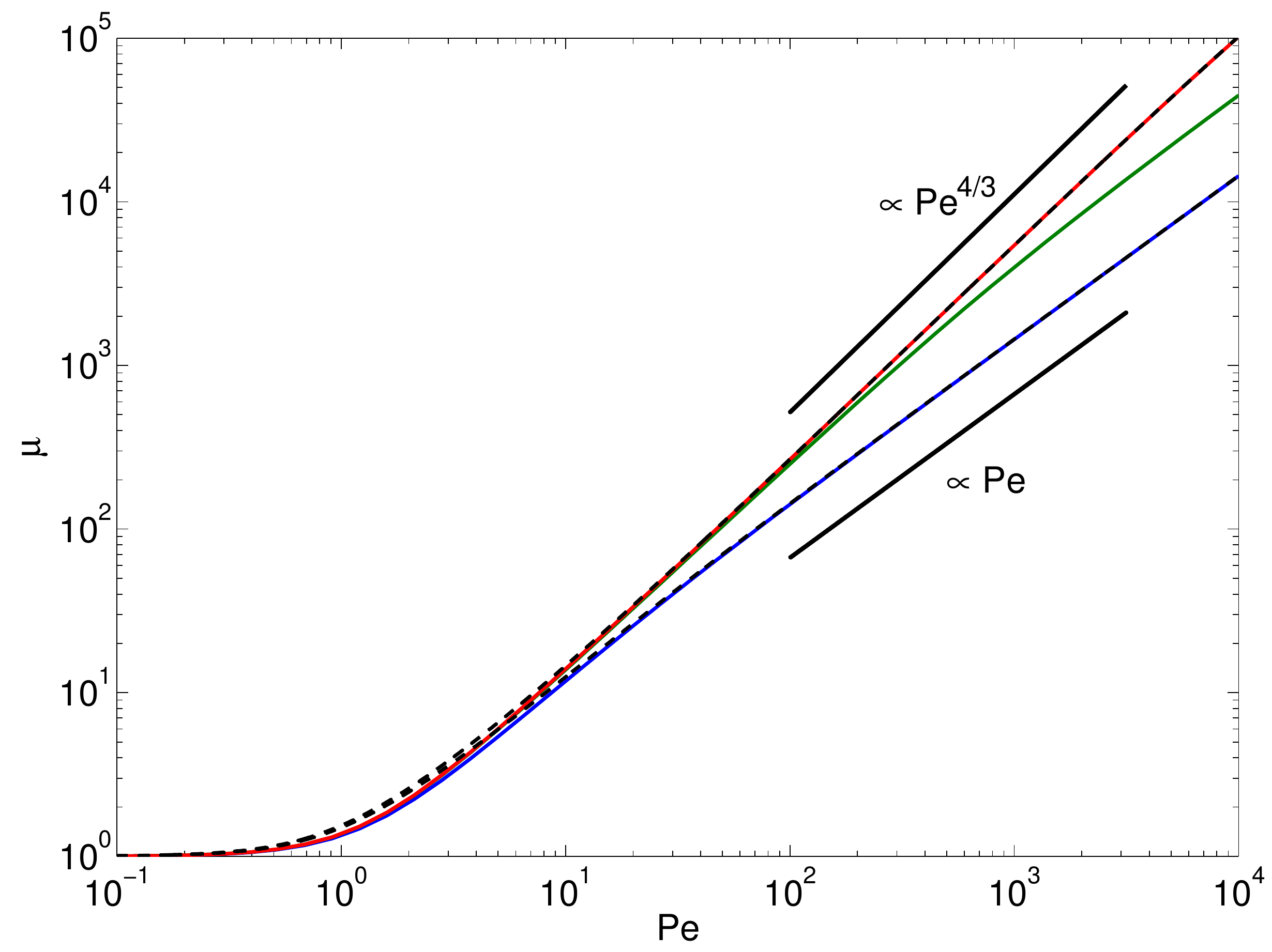}
\end{center}
\caption{Normalized stretching coefficient $\mu=2\lambda/3p$ as function of the P\'eclet number for $pL = 10^1$, $10^2$, and $10^3$ (bottom to top). In all cases, we set $l=1/p$, which corresponds to $L/l= pL$, and  $\Delta=1/3$. The dashed lines represent the solution of the asymptotic equation (\ref{eq:lagpar_flex}). The straight lines indicate the power-law dependencies $\mu \sim Pe^{4/3}$ for $L/l = 10^3$ and $\mu\sim Pe$ for $L/l = 10$ (cf. Eqs. (\ref{eq:limit_lagpar_pl}) and (\ref{eq:limit_lagpar_pe}), respectively)}  \label{fig:lagpar_n}
\end{figure}

{\em Flexible-Polymer Limit---}An analytical solution of Eq.~(\ref{eq:constraint_mode}) can easily be obtained for a flexible polymer, where $pL \gg 1$. In this case, the wavenumbers are given by $\zeta_n=n\pi/L$ and the eigenfunctions reduce to trigonometric functions \cite{harn:95} such that
\begin{align}
\int_{-L/2}^{L/2} \left( \frac{d \vphi_n(s)}{ds} \right)^2 ds \approx \zeta_n^2 \ .
\end{align}
Hence,  Eq.~(\ref{eq:constraint_mode}) turns into
%\begin{align}
%\sum_{n=1}^{\infty} \left( \frac{3}{\epsilon \zeta_n^2 + 2 \lambda} + \frac{v_0^2 l %\gamma^2}{k_BT(\epsilon \xi_n^2+2\lambda)[k_BT(\epsilon \xi_n^4+2\lambda\xi_n^2) +\gamma \gamma_R ]} %\right) = L \ .
%\end{align}
\begin{align}
\sum_{n=1}^{\infty} \left( \frac{3}{\epsilon \zeta_n^2 + 2 \lambda} + \frac{v_0^2 l \gamma^2}{k_BT(4 \lambda^2 k_BT +  \epsilon \gamma \gamma_R)\zeta_n^2 +  2\lambda \gamma \gamma_R   k_BT} \right) = L
\end{align}
including modes  up to order  $n^2$.
Evaluation of the sum yields
%\begin{align}
%& 1 = \frac{1}{\sqrt{\mu}}\coth(2pL\sqrt{\mu}) - \frac{1}{2pL\mu}
%  \\ \nonumber & +\frac{Pe^2}{8 \Delta }\left[  \sqrt{\frac{8}{9\mu}} \sqrt{\frac{1-\sqrt{1-\frac{2}{3\Delta l^3 p^3 \mu^2}}}{1-\frac{2}{3\Delta l^3 p^3 \mu^2}}}\coth\left( pL \sqrt{2\mu \left(1-\sqrt{1-\frac{2}{3\Delta l^3 p^3 \mu^2}}\right)} \right)  \\  & \left. - \sqrt{\frac{8}{9\mu}} \sqrt{\frac{1+\sqrt{1-\frac{2}{3\Delta l^3 p^3 \mu^2}}}{1-\frac{2}{3\Delta l^3 p^3 \mu^2}}}\coth\left( pL \sqrt{2\mu \left(1+\sqrt{1-\frac{2}{3\Delta l^3 p^3 \mu^2}}\right)} \right)    +  \frac{4}{3\sqrt{\mu}}\coth\left( 2pL\sqrt{\mu} \right) - \frac{2}{3pL\mu} \right] \ .
%\end{align}
\begin{equation}
\begin{split}
& \frac{3L\sqrt{2\lambda}\coth\left(L\sqrt{2\lambda/\epsilon}\right)-3 \sqrt{\epsilon}}{4\lambda \sqrt{\epsilon}} \\ &
+\frac{\gamma l v_0^2 L}{4\gamma_R k_BT \lambda }\left[\sqrt{\frac{2 \gamma \gamma_R \lambda }{ 4 k_BT \lambda^2 + \epsilon \gamma \gamma_R }} \coth\left(L \sqrt{\frac{2 \gamma \gamma_R \lambda }{4 k_BT \lambda^2 +  \epsilon \gamma \gamma_R}} \right)-\frac{1}{L} \right] = L ,
\end{split}
\end{equation}
or in terms of the P\'eclet number $Pe$ and $\Delta$ [Eq.~(\ref{eq:peclet})],
\begin{equation} \label{eq:lagpar_flex}
\begin{split}
& \frac{1}{\sqrt{\mu}}\coth\left(2pL\sqrt{\mu} \right)-\frac{1}{2pL\mu} \\ &
+ \frac{Pe^2}{6 \mu \Delta}\left[ \sqrt{\frac{\mu}{1+6 \mu^2 p^3 l^3 \Delta}} \coth\left( 2pL\sqrt{\frac{\mu}{1+6 \mu^2 p^3l^3 \Delta}} \right) - \frac{1}{2pL} \right] = 1 .
\end{split}
\end{equation}
The solution of this equation is compared with the exact solution of Eq.~(\ref{eq:constraint_mode}) in Fig.~\ref{fig:lagpar}. Evidently, we find good agreement for  $pL \gg 1$ and $Pe \gtrsim 10$. Taking into account modes of order $n^4$ or even $n^6$, leads to a better agrement between the results of the two equations.\\
Equation (\ref{eq:lagpar_flex}) yields the following asymptotic dependencies:
\begin{itemize}
\item For a passive polymer, $Pe =0$ implies $\mu=1$.
\item In the limit $pL \to \infty$ and $Pe < \infty$, i.e., $1 \ll \mu < \infty$,
\begin{align} \label{eq:limit_lagpar_pl}
\frac{1}{\sqrt{\mu}} + \frac{Pe^2}{ \mu^{3/2}  (6 p l \Delta)^{3/2}} = 1 .
\end{align}
Hence, in the asymptotic limit $pL \to \infty$, $\mu \sim Pe^{4/3}/pl$ (cf. Figs.~\ref{fig:lagpar} and \ref{fig:lagpar_n}). Note that when we set $l=1/p$, i.e., identify $l$ with the Kuhn length, $\mu$ is independent of the polymer length in the considered scaling regime. This is illustrated in Fig.~\ref{fig:lagpar_n}.
\item For $pL < \infty$ and $Pe \to \infty$, i.e., $\mu \gg 1$,
\begin{align} \label{eq:limit_lagpar_pe}
\frac{1}{\sqrt{\mu}} + \frac{Pe^2}{ \mu^{2}}  \frac{L}{54 p^2 l^3 \Delta^2} = 1 ,
\end{align}
 which yields  $\mu \sim Pe (L/l)^{3/2}/pL$ (cf. Fig~\ref{fig:lagpar_n}). Here, there remains a polymer-length dependence for $l=l_p$, namely $\mu \sim Pe \sqrt{pL}$.
\end{itemize}
In the asymptotic limit $Pe \to \infty$, we find a crossover of the Lagrangian multiplier from the power-law dependence $\mu \sim Pe^{4/3}$ to $\mu \sim Pe$.  In the latter regime, the Lagrangian multiplier  depends on polymer length. The crossover behavior is illustrated in Fig.~\ref{fig:lagpar_n}. The figure presents results for flexible polymers of various lengths, where the Kuhn segment length is identified with $l$, i.e., $pL= L/l$. The power-law dependence $\mu \sim Pe^{4/3}$ is specific to the large number of internal degrees of freedom of a polymer. This applies to flexible as well as semiflexible polymers. As is discussed in the next section, activity changes the properties of semiflexible polymers and they exhibit flexible polymer behavior at large P\'eclet numbers. However, in the asymptotic limit $Pe \to \infty$, activity causes a stretching of the polymer and a crossover to the dependence $\mu \sim Pe$ appears. The same relation is obtained for a finite-extensible active dumbbell, which lacks internal degrees of freedom \cite{wink:16}. Hence, the dynamical properties of active polymers are not only determined by the longest relaxation time, as is often the case for passive polymers, but the internal degrees of freedom play a much more significant role than for passive polymers.

\subsection{Relaxation Times}

The relaxation times [Eq.~(\ref{eq:relax_time})]
\begin{align} \label{eq:relax_time_mu}
\tau_n = \frac{\gamma}{3k_BT p} \left(\mu \zeta_n^2 + \frac{1}{4p^2} \zeta_n^4 \right)^{-1}
\end{align}
depend via $\mu$ on the activity $v_0$ (or $Pe$). We like to emphasize once more that this is a consequence of the finite extensibility of a polymer \cite{wink:16}. Neglecting this intrinsic property implies $\mu=1$ and the relaxation times are independent of the activity \cite{kais:15,sama:16}.  The presence of the factor $\mu$ gives rise to a particular dynamical behavior, specifically for semiflexible polymers.

In the limit of a flexible polymer, the relaxation times become
\begin{align} \label{eq:tau_asym_flex}
\tau_n = \frac{\gamma L^2}{3 \pi k_BT p} \frac{1}{\mu n^2} = \frac{\tau_R}{\mu n^2} \ ,
\end{align}
with the Rouse relaxation time $\tau_R = \gamma L^2/3 \pi k_B T p$ \cite{harn:95,doi:86}. Since, $\mu \geqslant 1$ is a monotonically increasing function of $Pe$, activity accelerates the relaxation process and the relaxation times become shorter. However, the mode-number dependence is not affected.

\begin{figure}[t]
\begin{center}
\includegraphics*[width=0.7\columnwidth]{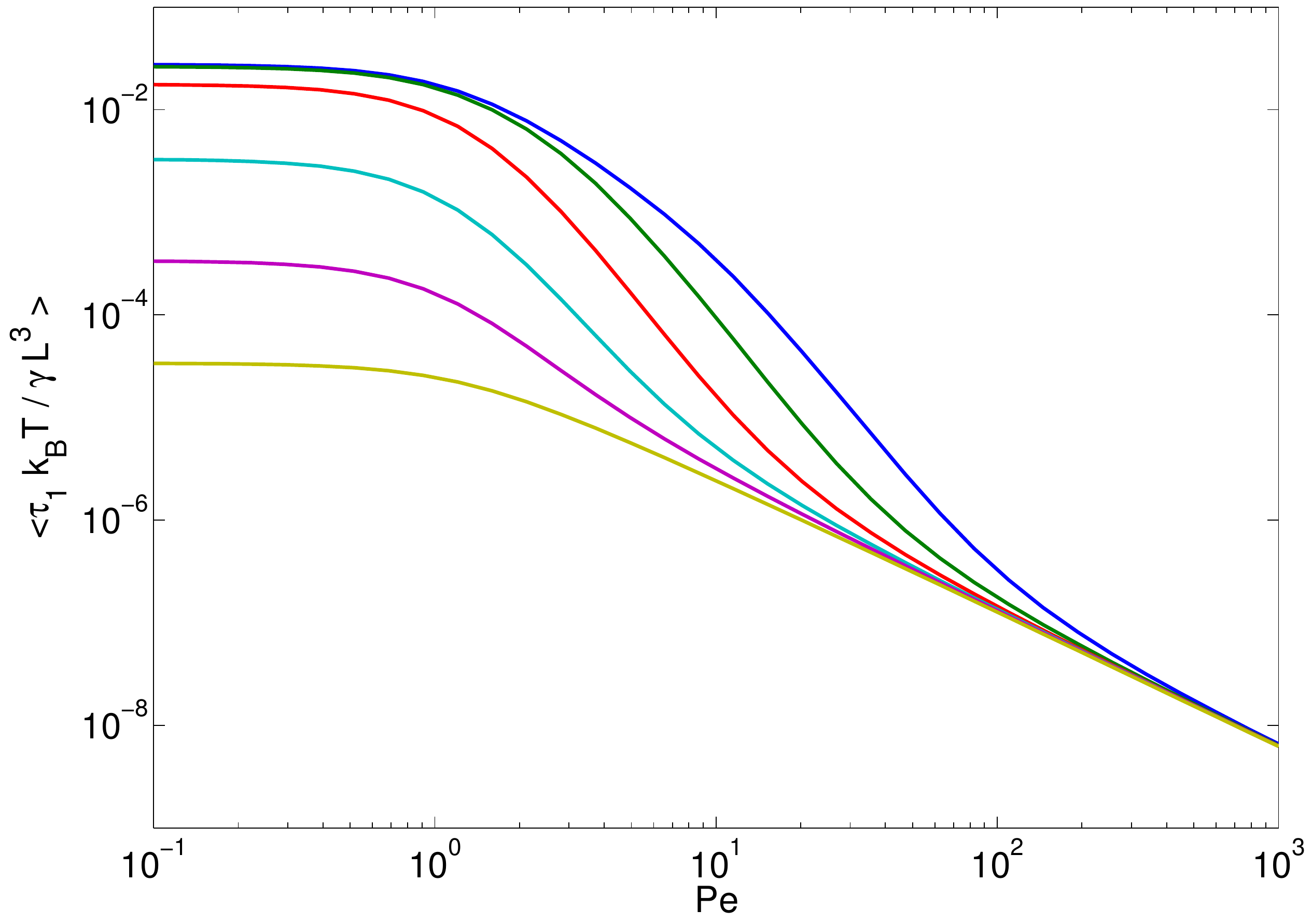}
\end{center}
\caption{Longest polymer relaxation times as function of the P\'eclet number for the bending stiffnesses ($L$ is fixed) $pL = L/2l_p= 10^3$, $10^2$, $10$, $1$, $10^{-1}$, and $10^{-2}$ (bottom to top). The other parameters are the same as in Fig.~\ref{fig:lagpar}. }  \label{fig:tau_1}
\end{figure}

The influence of activity on semiflexible polymers is much more substantial. For such polymers, $pL < 1$ and the $\zeta^4$-dependence (bending modes) typically  dominates the relaxation behavior.
However, with increasing activity, and hence $\mu$, the flexible modes ($\zeta_n^2$) in  Eq.~(\ref{eq:relax_time_mu}) dominate over the bending modes. Thus, the contribution $\mu \zeta_n^2$ determines the relaxation behavior of the polymer for $n^2 \lesssim 4 (pL)^2 \mu/\pi^2$. Only for larger modes, semiflexibility matters. As a consequence, starting from the large length-scale dynamics, activity induces a transition from semiflexible to flexible polymer behavior, which extends to smaller and smaller length scales with increasing $Pe$.
This behavior is illustrated in Fig.~\ref{fig:tau_1} for the longest polymer relaxation time $\tau_1$. For $pL \gg 1$, $\tau_1$ exhibits the predicted $1/\mu$ behavior [cf. Eq.~(\ref{eq:tau_asym_flex})], with $\tau_1 \sim Pe^{-4/3}$ for large  $Pe$. At $Pe \lesssim 1$, the relaxation times of the stiffer polymers are determined by the bending modes, and $\tau_1$ approaches the persistence-length and $Pe$ independent value
\begin{align} \label{eq:tau_1_rod}
\tau_1 = \frac{\gamma L^3}{36 k_B T}
\end{align}
with decreasing $pL$. The increase of $\mu$ with increasing P\'eclet number causes a decrease of the relaxation time $\tau_1$, and in the limit $Pe \gg 1$, the relaxation times assume the same asymptotic value of Eq.~(\ref{eq:relax_time}) independent of the stiffness.
Quantitatively, $\tau_1 \sim 1/\mu$ as soon as $\mu \gg (\pi/2pL)^{-2}$. The latter is already satisfied for rather moderate P\'eclet numbers on the order of $Pe \sim 10^1 - 10^2$.

\begin{figure}[t]
\begin{center}
\includegraphics*[width=0.7\columnwidth]{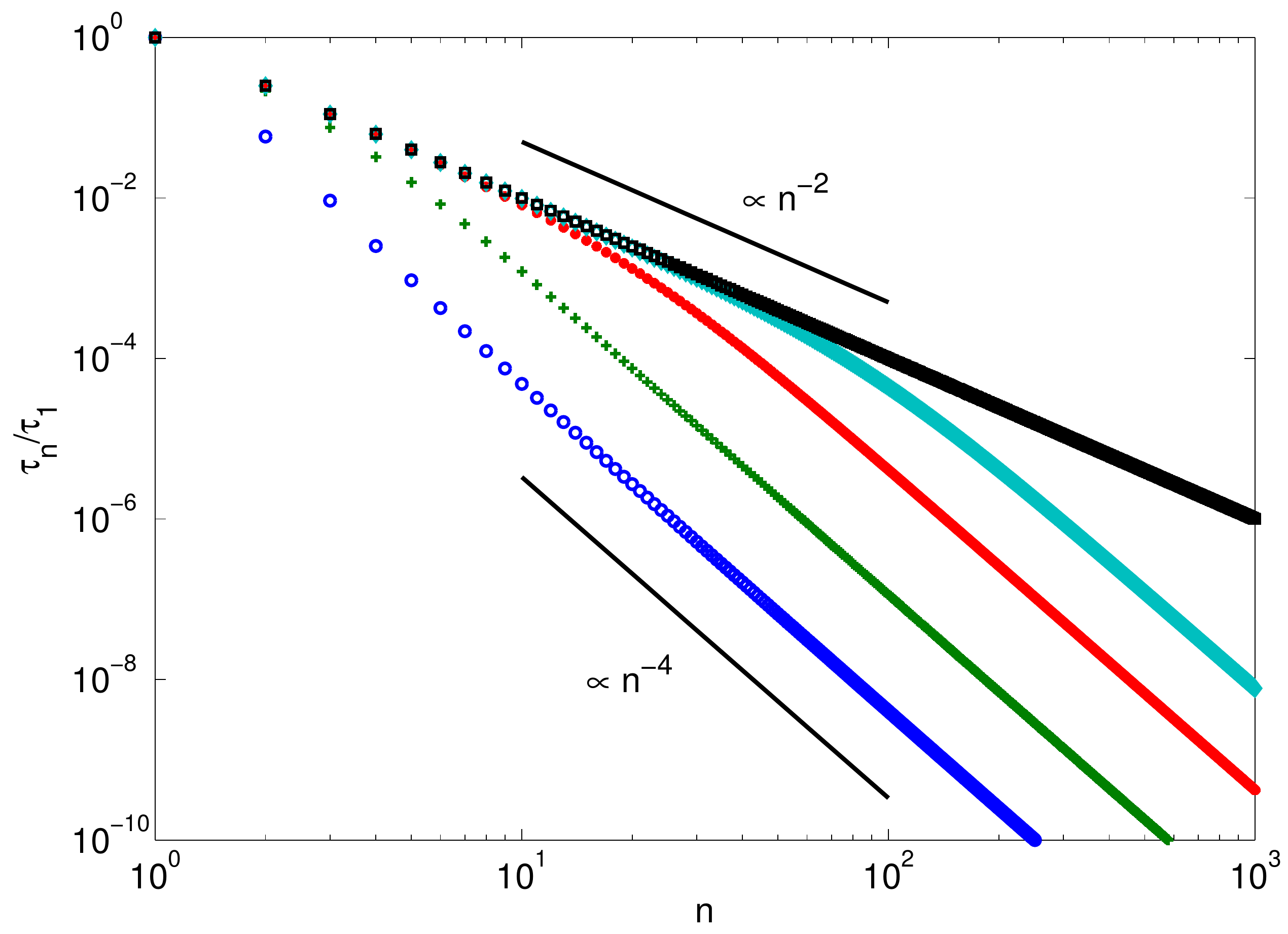}
\end{center}
\caption{Mode-number dependence of the relaxation times of active polymers with $pL=10^{-2}$ for the  P\'eclet numbers $Pe = 10^1$, $3 \times 10^1$,  $10^2$, and $5 \times 10^2$ (bottom to top). The black squares (top) show the mode-number dependence of a flexible polymer with $pL=10^3$. The other parameters are $L/l=10^3$ and $\Delta = 1/3$. The solid lines indicate the relations for flexible ($\sim n^{-2}$) and semiflexible ($\sim (2n-1)^{-4}$) polymers, respectively. $\tau_1$ is the longest relaxation time.}  \label{fig:tau_mode}
\end{figure}

Figure~\ref{fig:tau_mode} displays the dependence of the relaxation times $\tau_n$ of a stiff polymer on the mode number for various P\'eclet numbers. At low $Pe$, we find the well-know dependence $\tau_n/\tau_1 \sim (2n-1)^{-4}$ valid for semiflexible polymers \cite{wink:07.1,harn:95,arag:85}. With increasing $Pe$, the relaxation times increase, and for $Pe \gtrsim 50$ the small-mode-number relaxation times exhibit the dependence $\tau_n/\tau_1 \sim n^{-2}$ of flexible polymers.  At larger $n$, the relaxation times cross over to the semiflexible behavior again. However, the crossover point shifts to larger mode numbers with increasing activity. Taking the wavenumbers for flexible polymers, Eq.~(\ref{eq:relax_time_mu}) yields the condition $n> 2 pL\sqrt{\mu}/\pi$ for the dominace of bending modes. Hence, active polymers at large P\'eclet numbers appear flexible on large length and long time scales and only exhibit semiflexible behavior a small lengths scales.

\begin{figure}[t]
\begin{center}
\includegraphics*[width=0.7\columnwidth]{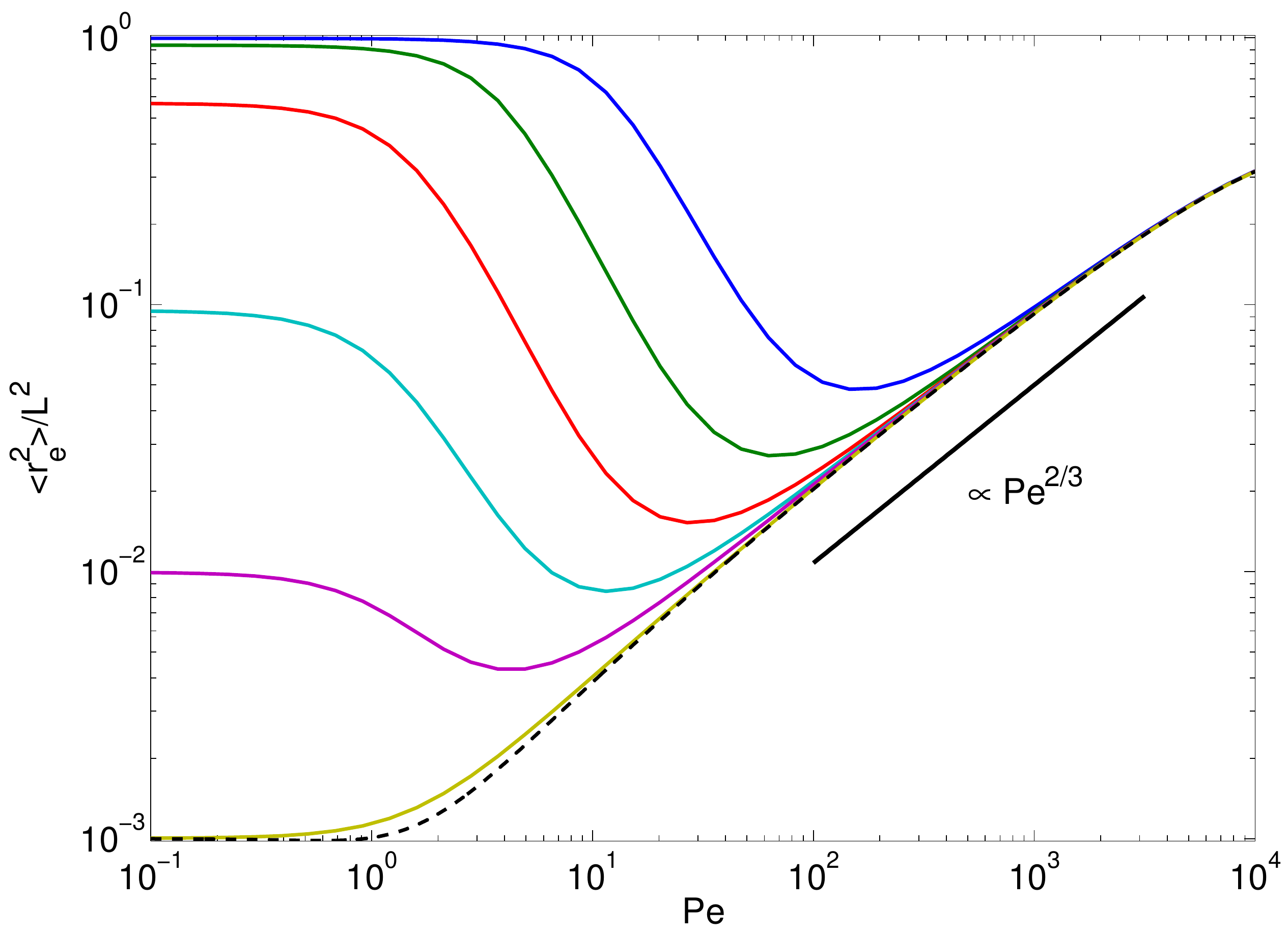}
\end{center}
\caption{Mean square end-to-end distances  as function of the P\'eclet number for the polymer bending stiffnesses $pL = 10^3$, $10^2$, $10$, $1$, $10^{-1}$, and $10^{-2}$ (bottom to top at $Pe=10^{-1}$). The other parameters are the same as in Fig.~\ref{fig:lagpar}. The dashed line represents the analytical solution of Eq.~(\ref{eq:end-to-end_flex}) with the Lagrangian multiplier of Eq.~(\ref{eq:lagpar_flex}).   }  \label{fig:end-to-end_distance}
\end{figure}

\subsection{Mean Square End-to-End Distance}

To characterize the conformational properties of the polymers, we consider the mean square end-to-end distance $\lla \bm r_e ^2\rra = \lla (\bm r(L/2) - \bm r(-L/2))^2 \rra$, which is given by
\begin{align} \label{eq:end-to-end}
\lla \bm r_e^2\rra = 4 \sum_{n=1}^\infty \lla \bs \chi_{2n-1}^2 \rra \vphi_{2n-1}^2(L/2)
\end{align}
in terms of the eigenfunction expansion (\ref{eq:eigen_expansion}), where
\begin{align} \label{eq:mode_amplitude}
\lla \bs \chi_n^2 \rra = \frac{3 k_BT}{\gamma} \tau_n + \frac{v_0^2l}{1+\gamma_R \tau_n} \tau_n^2 \ .
\end{align}
If the stretching coefficient $\lambda$ and, hence, the relaxation times were independent of the activity, the average mean square mode amplitudes (\ref{eq:mode_amplitude}) would increase quadratically with the P\'eclet number for $Pe \to \infty$ (cf. second term in the right-hand side of Eq.~(\ref{eq:mode_amplitude})). Thus, the mean square end-to-end distance would increase quadratically with $Pe$ \cite{kais:15}.
As shown in Fig.~\ref{fig:end-to-end_distance}, the constraint of a constant contour length drastically changes the activity dependence of the polymer conformations. In the limit of a flexible polymer (bottom curve of Fig.~\ref{fig:end-to-end_distance}), $\lla \bm r^2_e \rra$ increases with increasing P\'eclet number as $Pe^{2/3}$ from the passive equilibrium value $\lla \bm r^2_e \rra = L/p$. The mean square end-to-end distances of passive polymers itself increases with increasing persistence length, until the limit $\lla \bm r^2_e \rra = L^2$ is reached for $pL \to 0$. For bending stiffnesses $pL \lesssim 1$ and $Pe >1$, activity causes a significant shrinkage of the polymer over a wide range of P\'eclet numbers. Above a certain P\'eclet number, the actual value depends on the stiffness, the polymer swells again, but now similar to a flexible polymer and the asymptotic value  $\lla \bm r^2_e \rra = L^2/2$ is assumed for $Pe \rightarrow \infty$. This reflects the above mentioned activity-induced transition from semiflexible to flexible-polymer behavior.

\begin{figure}[t]
\begin{center}
\includegraphics*[width=0.6\columnwidth]{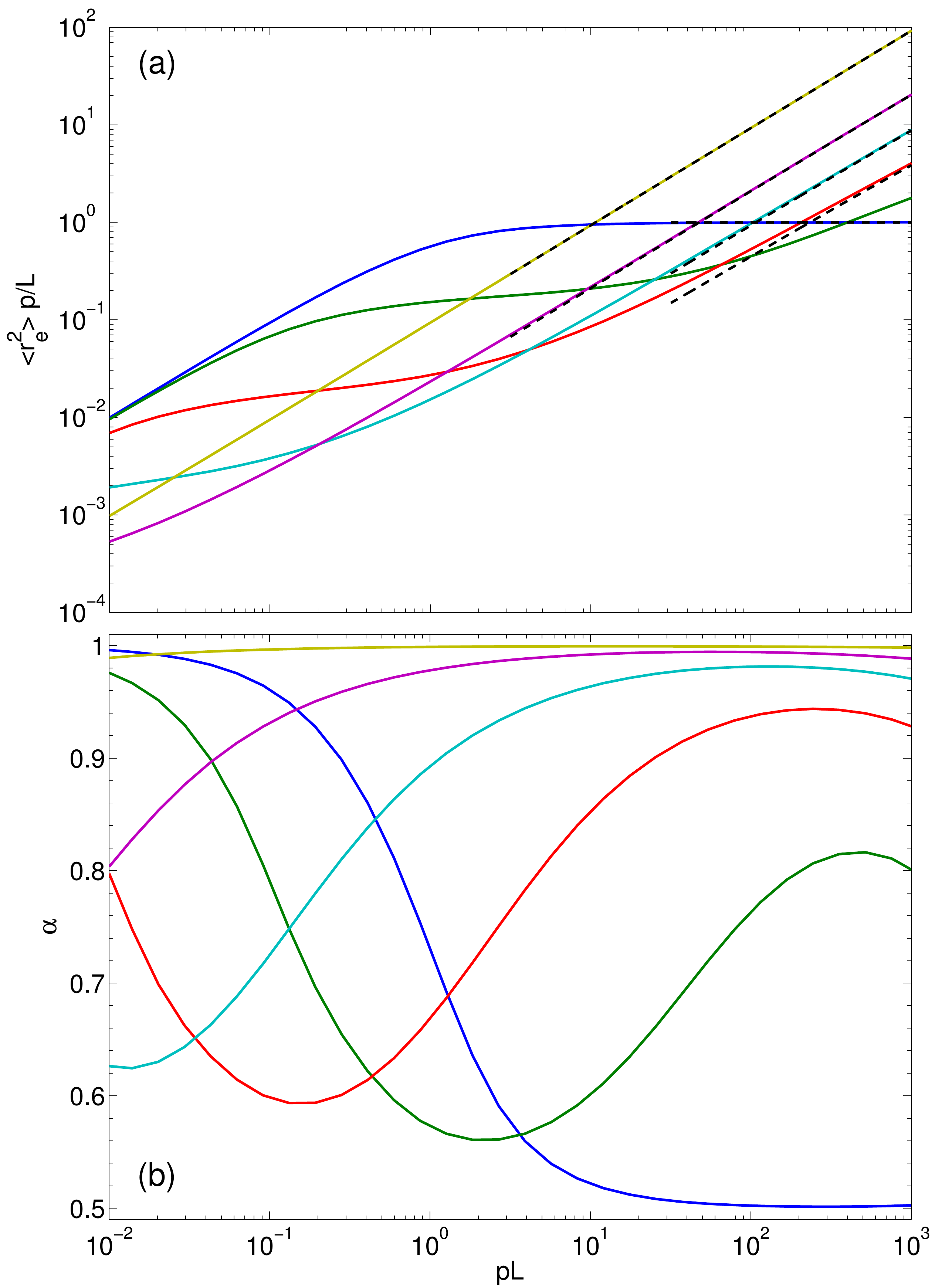}
\end{center}
\caption{(a) Mean square end-to-end distances and (b)  local slopes (Eq.~(\ref{eq:slope})) as function of the  polymer length ($pL$) for the P\'eclet numbers  $Pe = 0$, $3$, $10$, $30$, $10^{2}$, and $10^{3}$ (bottom to top at $pL=10^{3}$). The other parameters are the same as in Fig.~\ref{fig:lagpar}. The dashed lines in (a) represent the analytical solution of Eq.~(\ref{eq:end-to-end_flex}) with the Lagrangian multiplier of Eq.~(\ref{eq:lagpar_flex}).  } \label{fig:end-to-end_distance_pl}
\end{figure}

The scaling properties of $\lla \bm r^2_e \rra$ as function of polymer length ($pL$) are illustrated in Fig.~\ref{fig:end-to-end_distance_pl}(a). In addition,  Fig.~\ref{fig:end-to-end_distance_pl}(b) shows the local slope
\begin{align} \label{eq:slope}
\alpha = \frac{1}{2} \frac{d \log(\lla \bm r_e^2 \rra)}{ d \log(pL)} .
\end{align}
In the passive case $Pe=0$, $\lla \bm r^2_e\rra$ increases quadratically with increasing $pL$ for $pL < 1$  ($\alpha =1$, rodlike scaling). In the limit $pL \gg 1$, the flexible Gaussian polymer scaling  is obtained, where $\lla \bm r^2_e\rra = L/p$ ($\alpha = 1/2$), as is well know. In an active system, the local slope assumes the asymptotic value $\alpha =1$ for $pL \to 0$, independent of the P\'eclet number $Pe < \infty$. At a given $Pe >0$, the mean square end-to-end distance exhibits a monotonic progression with increasing $pL$ , but the local slope is non-monotonic. Starting from the asymptotic value $\alpha =1$, the local slope decreases first with increasing flexibility, i.e., $pL$, passes through a minimum, which depends on $Pe$, and increases again. This is illustrated in Fig.~\ref{fig:end-to-end_distance_pl}(b) for $Pe=3, \ 10,$ and $30$. The intermediate regime is rather broad, with local slopes almost as small as the value $1/2$ for simple Gaussian polymers.
In terms of scaling, we can identify a $pL$-regime for $pL> 1$, the actual range depends on $Pe$, where $\alpha$ is gradually increases with increasing P\'eclet number from the flexible polymer value $\alpha =1 /2$ to the rod limit $\alpha=1$. In addition, (smaller) scaling regimes exist in the crossover region, which shift to smaller $pL$ values with increasing $Pe$, with local slopes increasing from $\alpha =1/2$ with increasing P\'eclet number.
The slopes for $Pe \geqslant 3$ decrease for large $pL$ values. This is related to the selected density of active sites $N=L/l=10^3$ along the polymer. For $pL< 10^3$, a polymer is stiff on the length scale $p=1/l$. In contrary, for $pL >10^3$, the polymer becomes flexible on lengths scales smaller than $l$, which gives rise to the decrease of the local slope.

{\em Flexible-Polymer Behavior---}Evaluation of Eq.~(\ref{eq:end-to-end}) in the limit of flexible polymers taking into account modes up to $n^4$, but neglecting all $\epsilon$ terms, yields
\begin{align} \label{eq:end-to-end_flex}
\lla \bm r_e^2 \rra = \frac{L}{p \mu} + \frac{Pe^2 L}{6 p \mu \Delta} \left[
1 - \frac{\sqrt{1 + 6 p^3 l^3 \mu^2 \Delta }}{pL \sqrt{\mu}} \tanh \left( \frac{pL \sqrt{\mu}}{\sqrt{1 + 6 p^3 l^3 \mu^2 \Delta }} \right) \right] .
\end{align}
This equation exhibits the asymptotic behaviors:
\begin{itemize}
\item For finite $pL$ and $Pe \to \infty$, the argument of the hyperbolic tangent function becomes small and Taylor expansion gives
    \begin{align}
    \lla \bm r_e^2 \rra \approx \frac{Pe^2 L^3}{108 p^2 l^3 \Delta^2 \mu^2} .
    \end{align}
Insertion of the asymptotic behavior of Eq.~(\ref{eq:limit_lagpar_pe}) for the Lagrangian multiplier yields
$\lla \bm r_e^2 \rra \stackrel{Pe \to \infty}{\longrightarrow} L^2/2$.
Hence, the polymers assume nearly stretched conformations independent of the persistence length. This is visible in Fig.~\ref{fig:end-to-end_distance}.
\item  For $Pe\gg 1$, such that $1\ll \mu \ll \infty$, and $pL\to \infty$, the argument of the hyperbolic tangent function becomes large. By setting the hyperbolic tangent to unity, we obtain
    \begin{align}
    \lla \bm r_e^2 \rra \approx  \frac{L}{p \mu} \left(1 + \frac{Pe^2 }{6 \Delta}\right) .
    \end{align}
Insertion of the asymptotics  of Eq.~(\ref{eq:limit_lagpar_pl}) for the stretching coefficient yields
$\lla \bm r_e^2 \rra \approx  lL Pe^{2/3}$.
This dependence on the P\'eclet number is shown in Fig.~\ref{fig:end-to-end_distance} for the polymer with $pL =10^3$.
\end{itemize}

\section{Summary and Conclusions}

We have presented an analytical approach to study the conformational and dynamical properties of active semiflexible polymers. We have adopted a continuum representation of a polymer with a certain number of active segments. Each of the segments is considered as an active Brownian particle whose orientation changes independently in a diffusive manner. Alternatively, the active random process can be considered as an additional external correlated (colored) noise acting on the polymer \cite{elge:15,wink:16,sama:16,marc:16.1}. Active polymers have been considered before, both by theoretically and simulations \cite{live:01,sark:14,ghos:14,isel:15,sama:16,kais:15}. As an important extension of the previous studies, we have taken into account the finite polymer extensibility due to its finite contour length. As has been shown, this constraint changes the dynamical behavior of active dumbbells drastically \cite{wink:16}.  Taking into account the constraint by a Lagrangian multiplier leads to a linear equation, which is analytically tractable.

Evaluation of the polymer relaxation times shows a major influence of the finite contour length on the polymer dynamics. Models without such a constraint, e.g., the standard Rouse model \cite{doi:86}, would not be able to reproduce and capture the correct dynamics, as reflected in the strong dependence of the stretching coefficient (Lagrangian multiplier) on the P\'eclet number already for moderate $Pe$ values. In particular, the relaxation times decrease with increasing activity (P\'eclet number). Thereby, the influence of activity on stiff polymers is much more sever. Here, activity induces a transition from semiflexible polymer behavior, characterized by bending modes, to flexible polymer behavior, characterized by stretching modes, with increasing activity. Thereby, the affected length scale depends on the activity. For activities $Pe \gtrsim 20$, large length-scale and low-mode  number properties are altered. With increasing $Pe$, an increasing number of modes and hence smaller length scales are affected. Due to the continuous nature of the considered polymer model, the (very) small-scale properties will always be dominated by bending modes.

The effect on the relaxation times translates to the conformational properties.  In the simpler case of  flexible polymers, activity leads to a monotonous swelling of the polymers over a wide range of P\'eclet numbers in a power-law manner, which is dictated by the constraint. Hence, our theoretical prediction is very different from the relation $\lla \bm r_e^2\rra \sim Pe^2$ of a Rouse model derived in Ref.~\cite{kais:15} for any flexibility and P\'eclet number. For semiflexible polymers, with $pL \lesssim 10$, activity leads to shrinkage over a wide, stiffness-dependent range of P\'eclet numbers. At large $Pe$, the polymer conformations are comparable with those of flexible polymers.
An activity-induced shrinkage of semiflexible passive polymers embedded in a fluid of ABPs has been observed in simulations of two-dimensional systems \cite{shin:15,hard:14}, in qualitative agreement with our theoretical predictions. This supports the equivalence between intramolecular activity and the impact of external colored noise on the properties of semiflexible polymers (cf.  Sec.~\ref{sec:model}).

The simulation studies of Ref.~\cite{kais:15} for two-dimensional ABPO predict an activity induced shrinkage of self-avoiding polymers. These kind of shrinkage may be particular for 2D ABPS in combination with self-avoidance. As stated in Ref.~\cite{kais:15}, the polymer shrinkage at moderate P\'eclet numbers can be attributed activity-induced encaging by neighboring ABPs. The particular relevance of excluded-volume interactions in 2D systems
is also reflected in other studies, e.g., in Refs.~\cite{shin:15,isel:15,hard:14}. The activity-induced shrinkage of our 3D semiflexible polymers is of different origin. Here, self-avoidance does not play any role. In general,  self-avoidance is less important in 3D  than in 2D systems. Nevertheless, we expect interesting collective dynamical effects in 3D systems based on our studies of suspensions of 3D ABPs \cite{wyso:14}. Moreover,
the 2D simulations of Ref.~\cite{kais:15} suggest that the scaling relation of the mean square end-to-end distance with polymer length is unperturbed by the activity.  However, this should only apply to (very) small P\'eclet numbers, as is evident from Fig.~\ref{fig:end-to-end_distance_pl}, which suggest swelling of the polymer already for $Pe \gtrsim 1$ and an activity-induced modified scaling behavior for large $pL$ values.  Note that the P\'eclet number of Ref.~\cite{kais:15} is larger than ours due to the different definitions in terms of translational and rotational diffusion coefficient, respectively. We definitely find for $Pe > 10$ a wide crossover regime to the asymptotic scaling behavior of rodlike polymers, namely $\lla \bm r_e^2\rra \sim L^2$ (cf. Fig.~\ref{fig:end-to-end_distance_pl}).

Our studies illustrated the usefulness of basic polymer models for the understanding of the complex interplay between polymer entropy, stiffness, and activity. Extension of the current studies toward further dynamical properties and other propulsion preferences, e.g., along the tangent of the polymer contour, are under way.

Experimentally,  chains of ABPs can be synthesized by linearly connecting self-propelling Janus particles \cite{bech:16} by a flexible linker. A random distributed of linker sites on the colloid surface yields a random orientation of the propulsion directions of the individual ``monomers''. The ensemble average over various realizations corresponds to our description.


\section*{Acknowledgments}
Financial support by the Deutsche Forschungsgemeinschaft (DFG) within the priority program SPP 1726 “Microswimmers – from Single Particle Motion to Collective Behaviour” is gratefully acknowledged.

\renewcommand\bibname{References}


\begin{thebibliography}{-------}
\bibitem[com()]{comm1}
The {Eq.} (10) of {Ref.} 28 contains an error. {The} factor 2 in front of
  $v_0^2$ should ber replaced by unity.
\end{thebibliography}
\end{document}